# Video-rate holographic telepresence via single-shot, reference-free wavefront measurement

Minwook Kim[1,2,†], Chansuk Park[1,2,3,†], Chulmin Oh[1,2,†], KyeoReh Lee[1,2,4], Herve Hugonnet[1,2], YongKeun Park[1,2,5,*]

[1] Department of Physics, Korea Advanced Institute of Science and Technology (KAIST), 291, Daehak-ro, Yuseong-gu, Daejeon, 34141, Republic of Korea

[2] KAIST Institute for Health Science and Technology, KAIST, 291, Daehak-ro, Yuseong-gu, Daejeon, 34141, Republic of Korea

[3] Current affiliation: Samsung Electro-Mechanics, Maeyeong-ro 150, Yeongtong-gu, Suwon 16674, Republic of Korea

[4] Current address: Department of Applied Physics, Yale University, New Haven, 06520, Connecticut, USA

[5] Tomocube Inc., 141 Jukdong-ro, Yuseong-gu, Daejeon 34109, Republic of Korea

[†]These authors contributed equally.

[*]yk.park@kaist.ac.kr

**Abstract**: We present a reference-free holographic telepresence system that directly captures and replays complex optical wavefronts from a single intensity speckle measurement. Using a pre-characterized geometric phase diffuser, the incident field self-interferes to form a speckle pattern, from which the wavefront is recovered via a speckle-correlation scattering-matrix approach and refined using smoothed amplitude flow with Nesterov acceleration. The reconstructed phase is directly projected onto a spatial light modulator for holographic replay. We demonstrate volumetric refocusing, dynamic three-dimensional reconstruction, and sustained video-rate operation at approximately 28 frames per second with modest communication bandwidth. The results highlight measurement-driven wavefront acquisition as a practical pathway toward compact and physically faithful holographic telepresence.

# 1 Introduction

Telepresence aims to enable spatially separated users to share a realistic visual experience and a sense of co-presence by faithfully conveying three-dimensional (3D) scenes[1–3]. Since its early conceptualization in science fiction and virtual reality, telepresence has found practical applications in telemedicine, remote prototyping, advertising, immersive mapping, and entertainment[4–9]. Despite significant technological advances, achieving *physically faithful* telepresence—in which the optical behavior of a remote scene is preserved rather than visually approximated[10,11]—remains an open challenge.

A typical telepresence pipeline consists of imaging, streaming, processing, and display stages[12]. Most existing systems rely on depth cameras or multi-view imaging to capture scene geometry, followed by digital rendering to synthesize view-dependent images[13–15]. While such approaches benefit from commercial maturity and real-time operation, they fundamentally reconstruct *geometric representations* (e.g., depth maps, meshes, textures) rather than the underlying optical wavefront. As a result, the displayed views are visually plausible but not wavefront-faithful. Light-field and multi-view camera systems further encode directional information of light rays, improving parallax and viewpoint continuity[16,17], yet still require computational rendering or computer-generated holography and typically involve large data volumes, complex calibration, and substantial processing overhead[18]. More recently, data-driven methods combining neural scene representations—such as neural radiance fields (NeRF)—with holographic or incoherent imaging have demonstrated impressive visual fidelity for holographic communication[19–21]. These approaches infer view-dependent holograms from learned scene models and represent a powerful form of *rendered holography*. However, they remain computationally intensive and rely on synthetic approximations of optical propagation, rather than direct measurement of the complex optical field. Consequently, a fundamental gap persists between computationally synthesized telepresence and physically faithful wavefront reproduction.

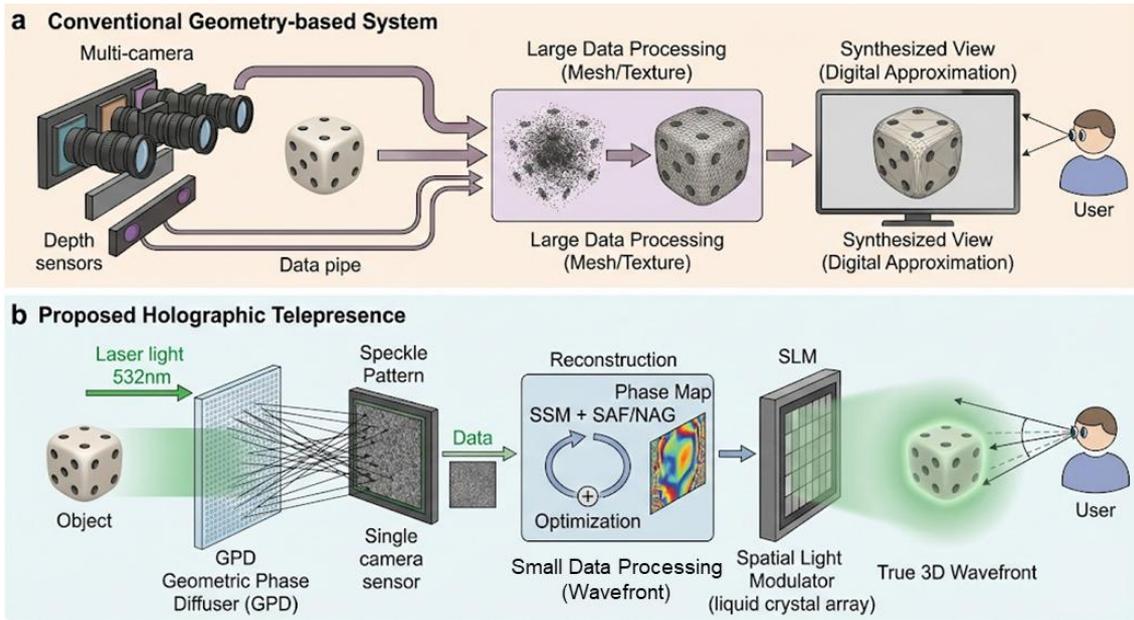

**Fig. 1 | Conceptual comparison between geometry-based telepresence and wavefront-based holographic telepresence.** (a) Conventional telepresence pipelines acquire scene geometry using multiple depth or light-field cameras. (b) Proposed holographic telepresence framework directly captures the complex optical wavefront of a three-dimensional scene from a single intensity-only speckle image using a pre-characterized geometric phase diffuser (GPD). The incident field is reconstructed via a reference-free speckle-correlation scattering-matrix (SSM) approach with iterative refinement, and the recovered phase is replayed on a phase-only spatial light modulator (SLM).

From an optical perspective, holographic telepresence ideally requires capture, transmission, and replay of the *complex optical wavefront* itself. Extensive work on wavefront control in complex and scattering media has established that transmission-matrix and speckle-correlation approaches enable deterministic reconstruction and replay of optical fields using spatial light modulators[22–24]. Conventional holographic imaging can, in principle, achieve this goal[25,26], but typically depends on interferometric reference beams and bulky, vibration-sensitive optical configurations that limit practicality for compact, scalable telepresence systems. Therefore, a key unmet need is a reference-free, data-efficient, and compact wavefront-based telepresence architecture capable of operating at video rates.

In this work, we address this gap by presenting a holographic telepresence system based on single-shot, reference-free speckle-correlation imaging and wavefront replay. By introducing a pre-characterized geometric phase diffuser between the object and the camera, the incident optical field self-interferes to form an intensity-only speckle pattern, from which the complex wavefront is recovered using a speckle-correlation scattering-matrix (SSM) approach[27–

[29]. The recovered field is refined through smoothed amplitude flow with Nesterov accelerated gradient optimization and directly projected onto a phase-only spatial light modulator for holographic replay. This end-to-end, measurement-driven pipeline enables volumetric refocusing and dynamic 3D holographic playback without explicit geometry estimation, multi-view acquisition, or interferometric reference beams.

We experimentally validate the proposed framework through volumetric refocusing, dynamic scene reconstruction, and system-level performance evaluation. The prototype demonstrates stable video-rate holographic telepresence at approximately 28 frames per second with modest communication bandwidth. Beyond immersive visualization, this work positions physically faithful wavefront telepresence as a missing link between computational scene representations and optical holographic displays.

## 2 Methods

### 2.1 Optical setup

The overall optical configuration of the holographic telepresence system consists of two functionally distinct stages: an imaging stage for single-shot wavefront acquisition and a display stage for holographic wavefront replay, as illustrated in Fig. 2.

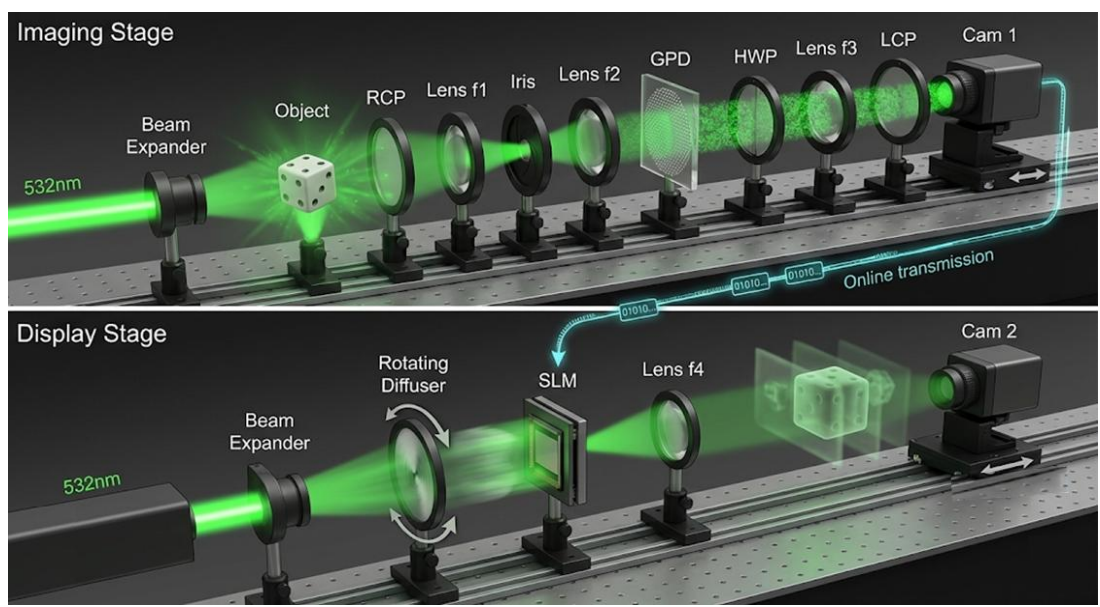

**Fig. 2 | Optical architecture of the holographic telepresence system.** The system consists of an imaging stage for

single-shot wavefront acquisition (top) and a display stage for holographic wavefront replay (bottom). RCP/LCP: right-/left-handed circular polarizer; HWP: half-wave plate. Focal lengths f1–f4 of lenses L1–L4 are 200 mm, 300 mm, 200 mm, and 400 mm, respectively.

2.1.1 Imaging stage: single-shot speckle-based wavefront acquisition

In the imaging stage, a spatially coherent laser source (532 nm, Samba, Cobolt AB) illuminates the object. The scattered optical field is collected and spectrally band-limited at the Fourier plane of lens L1 by a circular iris diaphragm (radius 0.6 mm), which defines the system's spatial frequency support and controls oversampling.

The filtered field is relayed by lens L2 onto a pre-characterized geometric phase diffuser (GPD; Thorlabs, Inc.), which acts as a known random phase mask. The diffuser induces self-interference of the incident field, generating an intensity-only speckle pattern at the sensor plane without the need for an external reference beam. Circular polarization optics—comprising a right-handed circular polarizer (RCP), a half-wave plate (HWP), and a left-handed circular polarizer (LCP)—are employed to isolate the geometric-phase modulation, enabling a compact and reference-free optical configuration.

The GPD has a pixel pitch of 30 μm with a resolution of 512 × 512 pixels. Because the phase profile of the diffuser is known a priori. The GPD rotation and magnification mismatches between the physical and numerical models were refined, yielding an optimized rotation of 0.667° and a magnification factor of 0.999. After propagation through lens L3, the resulting speckle intensity is recorded by a monochrome camera (Cam1, MD120MU-SY, XIMEA GmbH) with a pixel size of 3.1 μm. Owing to the focal length ratio between L2 and L3, the effective GPD pixel size at the reconstruction plane is 20 μm.

2.1.2 Display stage: holographic wavefront replay

In the display stage, the numerically reconstructed phase distribution is loaded onto a phase-only spatial light modulator (SLM; LC2012, Holoeye; pixel pitch 36 μm). To suppress coherent speckle artifacts during optical replay, a rotating diffuser is placed upstream of the SLM. The modulated wavefront is then projected by lens L4 to form the three-dimensional holographic image in free space.

For experimental validation of the projected wavefront, a second camera (Cam2; pixel size 4.8 μm, Flea3 FL3-U3-13Y3M, FLIR) is mounted on a motorized translation stage. The camera position is scanned along the optic axis for depth-resolved verification of volumetric wavefront reply.

**2.2 Reconstruction algorithm**

To recover the complex optical field at the iris plane from a single-shot, intensity-only speckle measurement, we adopt a two-stage reconstruction framework that combines the speckle-correlation scattering matrix (SSM) method with iterative refinement using Smoothed Amplitude Flow (SAF)[27] and Nesterov accelerated gradient (NAG) optimization[30,31]. The SSM method provides a physics-based initial estimate of the complex field, while the subsequent SAF + NAG refinement improves reconstruction fidelity and temporal stability, which are essential for dynamic holographic telepresence.

2.2.1 Speckle-correlation scattering matrix (SSM)

The SSM method reconstructs both the phase and amplitude of an optical field from a single intensity speckle image by exploiting the statistical correlations inherent to speckle patterns[27]. Spatio-temporal correlations of complex optical channels play a critical role in maintaining reconstruction fidelity under dynamic conditions[32–34]. Unlike conventional holographic approaches that rely on external reference beams or interferometric configurations, SSM enables reference-free complex-field recovery using a compact optical setup. In the proposed system, the SSM algorithm reconstructs the complex field at the iris plane, which serves as a well-defined intermediate representation of the incident wavefront. The object-plane wavefront is subsequently obtained by applying a two-dimensional Fourier transform to the reconstructed iris-plane field, consistent with the system's Fourier-optical geometry. This SSM-based initialization provides a reliable starting point for iterative refinement, mitigating convergence to poor local minima in subsequent phase-retrieval optimization. Detailed mathematical derivations and implementation procedures of the SSM method are provided below.

2.2.2. Iteration combining SAF and NAG algorithms

The smoothed amplitude flow (SAF) algorithm is employed to refine the initial complex-field estimate by minimizing a smoothed amplitude-based discrepancy between the predicted and measured speckle fields[30]. Phase-retrieval problems are fundamentally non-convex, motivating algorithmic strategies that emphasize stable convergence and robustness over strict optimality[35]. Recent studies have formulated computer-generated holography as a non-convex inverse problem, highlighting the central role of optimization strategies that balance convergence stability, reconstruction fidelity, and computational efficiency[36]. Owing to the non-convex nature of the underlying loss function, reliable convergence requires a physically meaningful initialization, which is provided by the SSM reconstruction described above.

To accelerate convergence and improve numerical robustness, we incorporate Nesterov accelerated gradient (NAG) optimization[31] into the SAF framework. This momentum-based update scheme enhances convergence speed and, more importantly, suppresses frame-to-frame fluctuations during iterative reconstruction, which is critical for temporally stable holographic replay of dynamic scenes (see Results 3.2).

For benchmarking purposes, we also evaluated the Wirtinger Flow (WF) algorithm, a widely used phase-retrieval approach for complex-field reconstruction[37]. To ensure a fair comparison, both WF and SAF + NAG were initialized with the same SSM-derived complex field.

2.2.3 Notation

We model the formation of the measured speckle field as a known linear mapping between the incident optical field and the sensor plane. Let $\mathbf{x} \in \mathbb{C}^N$ denote the complex-valued incident field expressed in an input-mode basis, and let $\mathbf{T} \in \mathbb{C}^{M \times N}$ denote the known transmission matrix that maps the input modes to the sensor pixels[38,39]. The resulting complex speckle field at the measurement plane is given by

$$\boldsymbol{y} = \sum_{a=1}^{N} x_a t_a = \mathbf{T}\mathbf{x}, \qquad (1)$$

where $\mathbf{T} = [t_1, \ldots, t_N]$ and $t_a$ represents the response of the $a$-th input mode at the sensor plane. The camera records only the intensity of the complex field,

$$I(r) = |y(r)|^2, \ r \in \{1, \dots, M\}, \tag{2}$$

where $r$ indexes the sensor pixels. Spatial averaging over the sensor coordinates is denoted by

$$\langle \cdot \rangle_r = \frac{1}{M} \sum_{r=1}^{M} (\cdot). \tag{3}$$

The oversampling ratio, denoted by $\gamma$, is defined as the ratio between the number of measured output modes and the number of unknown input modes, $\gamma = N_{out}/N_{in}$, where $N_{out}$ corresponds to the number of independent camera sampling modes and $N_{in}$ represents the number of resolvable input modes of the incident optical field. In our system, $N_{out}$ is determined by the effective camera sampling area after cropping, while $N_{in}$ is set by the maximum number of spatial modes supported by the diffuser field of view and the optical magnification. As $\gamma$ increases, higher-order correlation terms in the speckle statistics decay, enabling the rank-one approximation that underlies the effectiveness of the SSM reconstruction. The resulting oversampling ratio $\gamma > 1$ ensures sufficient redundancy for reliable speckle-correlation-based phase and amplitude retrieval.

2.2.4 SSM Calculation via power iteration

The SSM method reconstructs the incident field by constructing a speckle-correlation scattering matrix from second-order intensity statistics. Using the notation introduced above, the correlation matrix $\mathbf{Z} \in \mathbb{C}^{N \times N}$ is defined as

$$Z_{pq} = \frac{1}{\Sigma_p \Sigma_q} \left[ \langle t_p^* t_q y^* y \rangle_r - \langle t_p^* t_q \rangle_r \langle y^* y \rangle_r \right], \tag{4}$$

where $t_p$ denotes the sensor-plane response of the $p$-th input mode, $\Sigma_p = \langle |t_p|^2 \rangle_r$.

Under the assumption of a random (diffuse) optical field, application of Wick's theorem[40] yields

$$Z_{pq} = \alpha_p \alpha_q^* + \frac{1}{\Sigma_p \Sigma_q} \langle t_p^* y^* \rangle_r \langle t_q y \rangle_r, \tag{5}$$

where the vector $\alpha$ is proportional to the incident field coefficients $x_a$.

As the oversampling ratio $\gamma$ increases, the second term in Eq. (5) decays, and the $\mathbf{Z}$ becomes approximately rank-one. Consequently, the principal eigenvector of $\mathbf{Z}$ converges to $\alpha$, and hence to the incident field $\mathbf{x}$ up to a global phase factor.

Direct construction and eigenmode decomposition of $\mathbf{Z} \in \mathbb{C}^{N \times N}$ are computationally prohibitive. Instead, we recover the dominant eigenvector using power iteration, which requires only repeated evaluation of the matrix–vector product $\mathbf{Z}\mathbf{v}$.

Let $t_p(r)$ denote the $r$-th element of the $p$-th transmission vector, and define the intensity fluctuation $\Delta I(r) = I(r) - \langle I \rangle_r$. Using the identity $y^* y = I$, Eq. (5) can be rewritten as

$$Z_{pq} = \frac{1}{\Sigma_p \Sigma_q} \langle t_p^* t_q \Delta I \rangle_r . \tag{6}$$

Expanding the spatial average yields

$$\langle t_p^* t_q \Delta I \rangle_r = \frac{1}{M} \sum_{r=1}^{M} t_p^*(r) \Delta I(r) t_q(r) = \left( \mathbf{T}^\dagger \mathrm{diag}(\Delta I) \mathbf{T} \right)_{pq}, \tag{7}$$

from which the correlation matrix can be expressed compactly as

$$Z = \frac{1}{M} \mathbf{D}^{-1} \mathbf{T}^\dagger \mathrm{diag}(\Delta I) \mathbf{T} \mathbf{D}^{-1}, \tag{8}$$

where $\mathbf{D} = \mathrm{diag}(\Sigma_1, \ldots, \Sigma_N)$, $\Sigma_p = \langle |t_p(r)|^2 \rangle_r$, represents the average energy of the $p$-th transmission mode at the sensor plane, and the diagonal normalization by $\mathbf{D}^{-1}$ removes mode-dependent power variations.

For an arbitrary vector $\mathbf{v} \in \mathbb{C}^N$, the action of $\mathbf{Z}$ is therefore given by

$$\mathbf{Z}\mathbf{v} = \frac{1}{M} \mathbf{U}^\dagger [\Delta I \odot (\mathbf{U}\mathbf{v})], \tag{9}$$

where $\mathbf{U} = \mathbf{T} \mathbf{D}^{-1}$ and $\odot$ denotes element-wise multiplication.

Explicit construction of $\mathbf{U} \in \mathbb{C}^{N \times N}$ is avoided by exploiting the fact that, in our optical configuration, the transmission operator is implemented via Fourier propagation and a known geometric phase diffuser. Accordingly, we define the forward operator

$$\mathcal{O}(\mathbf{X}) = \frac{1}{M} \mathcal{F}^{-1}(\mathcal{F}(\mathbf{X}) \odot \mathbf{G}), \tag{10}$$

where $\mathcal{F}(\cdot)$ denotes the Fourier transform and $\mathbf{G}$ is the known transmission function of the geometric phase diffuser. The corresponding adjoint operator $\mathcal{O}^\dagger(\cdot)$ is used for back-propagation in both the SSM power iteration and the subsequent SAF refinement. The complete power-iteration procedure for SSM initialization is summarized in Algorithm 1.

**Algorithm 1: Speckle-correlation scattering matrix (SSM) reconstruction via power iteration**

Input: Measured speckle intensity $I_{meas}$, Initial random complex field estimate $\mathbf{x}_{SSM}^{(0)}$

Output: Initial object field estimate $\mathbf{x}_{SSM}$

1. Normalize
$$\mathbf{x}_{SSM}^{(0)} \leftarrow \mathbf{x}_{SSM}^{(0)}/\left|\mathbf{x}_{SSM}^{(0)}\right|$$

2. Compute intensity fluctuation
$$\Delta I = I_{meas} - \langle I_{meas}\rangle_r$$

3. **for** $k = 1$ to $K_{SSM}$ **do**

   Power iteration for principal eigenvector:

   4. Update
$$\mathbf{x}_{SSM}^{(k)} \leftarrow \mathcal{O}^{\dagger}\left(\Delta I \odot \mathcal{O}\left(\mathbf{x}_{SSM}^{(k-1)}\right)\right)$$

   5. Normalize
$$\mathbf{x}_{SSM}^{(k)} \leftarrow \mathbf{x}_{SSM}^{(k)}/\left|\mathbf{x}_{SSM}^{(k)}\right|$$

6. **end**

7. Return $\mathbf{x}_{SSM}^{(K_{SSM})}$

In this work, we set $K_{SSM} = 100$, which was sufficient to ensure stable convergence in all experiments. While the SSM provides a fast and direct initialization of the complex field, residual noise arising from the higher-order correlation term in Eq. (5) remains, motivating subsequent iterative refinement.

2.2.5 Smoothed amplitude flow (SAF)

The smoothed amplitude flow (SAF) algorithm is employed to refine the initial complex-field estimate obtained from the SSM reconstruction by minimizing a smoothed amplitude-based discrepancy between the predicted and measured speckle fields[30]. Unlike intensity-based least-squares formulations, SAF introduces smoothing parameter and vector that regularizes the amplitude mismatch, improving numerical stability in the presence of measurement noise.

The measured amplitude is given by $\psi = \sqrt{I_{meas}}$. SAF minimizes the smoothed amplitude loss function

$$\mathcal{L}_{SAF}(\mathbf{x}) = \frac{1}{2}\left\|\sqrt{|\mathbf{u}|^2 + \epsilon^2} - \sqrt{\psi^2 + \epsilon^2}\right\|_2^2, \tag{11}$$

where $\epsilon$ is a smoothing vector that prevents gradient singularities when the predicted amplitude approaches zero. In this paper, we implicitly set the smoothing parameter to 2 and the smoothing vector to $\epsilon = \psi$.

Because the SAF objective is non-convex, convergence critically depends on a physically meaningful initialization. In this work, the SSM-derived field estimate provides such an initialization, enabling SAF to converge reliably toward a consistent solution rather than becoming trapped in poor local minima. This combination leverages the complementary strengths of both methods: SSM provides a fast, physics-based approximation of the incident field, while SAF suppresses residual noise originating from higher-order speckle correlations. Gradient computation and iterative updates are performed using the adjoint operator $\mathbf{T}^\dagger$, consistent with the forward optical model described above. The SAF cost function was optimized with Nesterov momentum.

Unless otherwise specified, all operations, including division and the element-wise product $\odot$, are performed element-wise. The operator $\angle(\cdot)$ denotes extraction of the phase. The combined SAF + NAG update scheme is detailed in Algorithm 2.

---

**Algorithm 2: SAF + NAG implementation**

1. Initialize $\mathbf{x}^{(0)} = \mathbf{x}_{SSM}$. Set $\delta = 10^{-8}$ and $t_0 = 1$.
2. **for** $k = 1$ to $K$ **do:**
    3. Forward projection:
    $$\mathbf{y}^{(k)} = \mathcal{O}(\mathbf{x}^{(k)})$$
    4. Weight field:
    $$\mathbf{w}^{(k)} = \frac{|\mathbf{y}^{(k)}|}{\psi + \delta}$$
    5. Phase-corrected field:
    $$\mathbf{p}^{(k)} = \left(\sqrt{(\mathbf{w}^{(k)})^2 + 1} - \sqrt{2}\right) \odot \frac{\mathbf{w}^{(k)}}{\sqrt{(\mathbf{w}^{(k)})^2 + 1}} \odot \psi \odot e^{i \cdot \angle \mathbf{y}^{(k)}}$$
    6. Gradient (back-projection):
    $$\mathbf{g}^{(k)} = \mathcal{O}^\dagger(\mathbf{p}^{(k)})$$
    7. NAG update:
    $$\mathbf{z}^{(k)} = \mathbf{x}^{(k)} - \eta \odot \mathbf{g}^{(k)}$$
    $$t^{(k+1)} = \frac{1 + \sqrt{1 + 4(t^{(k)})^2}}{2}$$
    $$\mathbf{x}^{(k+1)} = \mathbf{z}^{(k)} + \left(\frac{t^{(k)} - 1}{t^{(k+1)}}\right)(\mathbf{z}^{(k)} - \mathbf{z}^{(k-1)})$$
8. **end**
9. Return $\mathbf{x}^{(K)}$

---

Here, $\mathbf{x}^{(0)}$ denotes the initial field estimate obtained from the SSM reconstruction, and $\mathbf{x}^{(k)}$ represents the output of the iterative refinement at iteration $k$. The vector $\psi$ corresponds to the square root of the measured speckle intensity recorded by the camera. The step size $\eta$ is defined as $\eta = \sqrt{\sum_{i=1}^{M} \psi_i^2}/max(|x^{(0)}|)$, which adaptively scales the gradient update according to the measured signal energy. To avoid numerical instability due to division by zero, a small constant $\delta = 10^{-8}$ is introduced. The initial Nesterov momentum parameter is set to $t_0 = 1$. The total number of iterations was set to $K = 100$, which was sufficient to ensure convergence in all experiments, as determined by a relative update criterion $\|x^{(k)} - x^{(k-1)}\|/\|x^{(k-1)}\| < 10^{-2}$.

## 3. Results

### 3.1 Volumetric refocusing

We first validate the physical fidelity of the proposed holographic telepresence system by examining its ability to reconstruct and refocus a three-dimensional scene from a single-shot speckle measurement. As a test object, three dice were positioned at two distinct axial planes, with two dice placed at $z = 0$ mm and one die at $z = 250$ mm. Reference images of the scene were captured using a conventional camera for comparison (top row of Fig. 3(c)). Figure 3(a) shows the raw speckle intensity image recorded at the sensor plane through the geometric phase diffuser, and Fig. 3(b) presents the reconstructed complex optical field at the iris plane together with the corresponding object-plane field obtained via Fourier transformation. Using the recovered complex wavefront, numerical refocusing was performed by propagating the field to three axial distances, $z_1 = 0$ mm, $z_2 = 125$ mm, and $z_3 = 250$ mm (middle row of Fig. 3(c)).

At $z_1$, the two near dice appear sharply focused while the distant die is blurred, whereas at $z_3$ the far die is in focus and the near dice are defocused. At the intermediate plane $z_2$, all objects appear blurred. These depth-dependent focus transitions confirm that the reconstructed field preserves the correct axial phase relationships required for volumetric wavefront propagation, rather than representing a depth-layered intensity approximation. Propagation-adaptive holographic frameworks have demonstrated that physically consistent three-dimensional image formation requires explicit treatment of optical propagation across depth, rather than single-plane optimization[41].

To further validate that the recovered wavefront can be physically replayed, the numerically reconstructed phase was projected on the spatial light modulator and physical light propagation in free space. The replayed hologram was recorded using a second camera mounted on a translation stage, with the camera focus adjusted to match the same axial positions $z_1$, $z_2$, and $z_3$ (bottom row of Fig. 3(c)). The optically replayed images exhibit depth-dependent focus behavior that closely matches both the numerical refocusing results and the conventional reference images. The agreement between numerical refocusing and optical replay demonstrates that the proposed system captures a physically consistent three-dimensional wavefront from a single intensity-only measurement and enables faithful volumetric reconstruction and holographic playback. A corresponding dynamic demonstration is provided in Supplementary Movie 1.

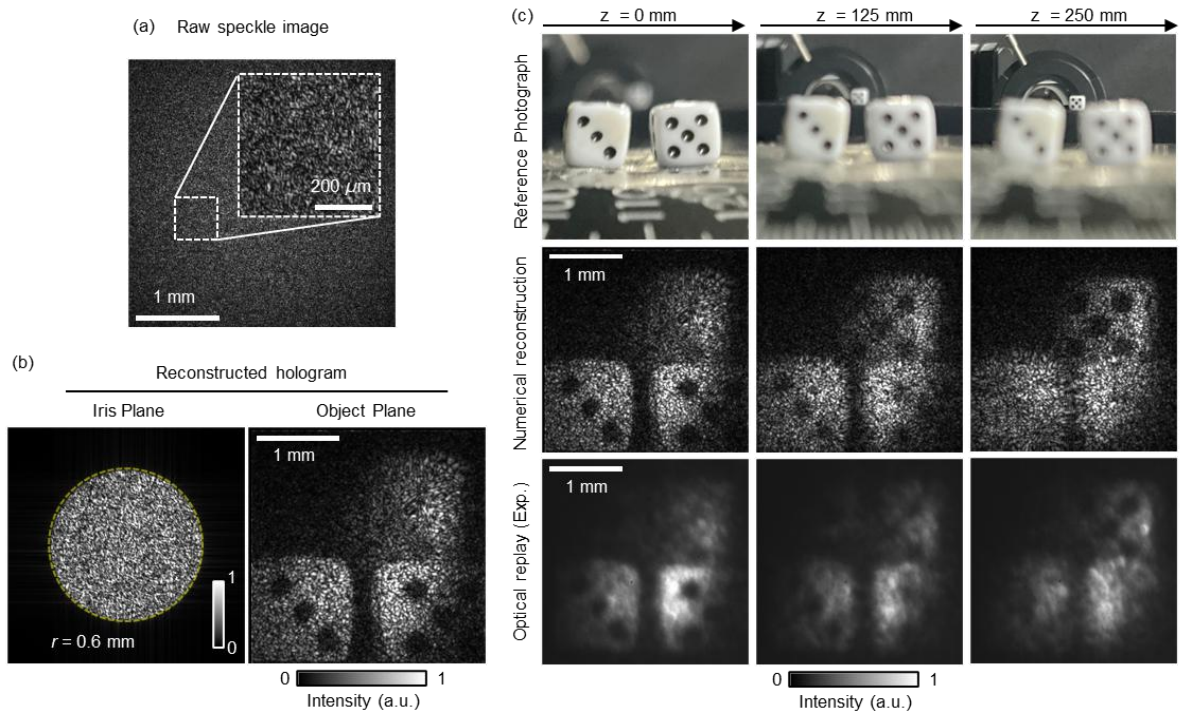

**Fig. 3 | Volumetric wavefront reconstruction and optical replay of a three-dimensional scene.** (a) Single-shot speckle intensity image captured at the camera plane through the geometric phase diffuser. (b) Reconstructed complex optical field at the iris plane (left) and corresponding object-plane field (right), obtained by applying a two-dimensional Fourier transform to the iris-plane field. The circular mask indicates the effective iris aperture. (c) Depth-resolved validation of volumetric reconstruction. Top row: reference images of three dice captured with a conventional camera, with two dice positioned at $z_1 = 0$ mm and one die at $z_3 = 250$ mm. Middle row: numerically refocused intensity images of the reconstructed wavefront at propagation distances $z_1 = 0$ mm, $z_2 = 125$ mm, and $z_3 = 250$ mm, respectively. Bottom row: experimentally replayed holograms recorded by a second camera during SLM-based wavefront projection, with the camera focus translated to match each depth plane.

**3.2 Algorithmic stability test**

To assess the temporal robustness of the reconstruction algorithms under dynamic conditions, we performed a comparative analysis between Wirtinger Flow (WF) and the proposed SAF + NAG solver. Figure 4 shows representative frames sampled every 30 frames from a 150-frame video sequence reconstructed using each method. For the static scene, in which the three dice remain stationary [Fig. 4(a,b)], both WF and SAF + NAG yield comparable reconstruction quality with similar contrast and negligible temporal variation across frames. This result indicates that both algorithms are capable of reliably reconstructing stationary objects from single-shot measurements.

In contrast, clear differences emerge under dynamic conditions, where the dice undergo lateral translation [Fig. 4(c,d)]. Reconstructions obtained with WF exhibit pronounced frame-to-frame flicker and non-uniform brightness, particularly at the beginning and end of the sequence (frames 0 and 150). These artifacts indicate a lack of temporal consistency, reflecting the sensitivity of WF to small inter-frame variations during iterative optimization. Comparative studies have shown that conventional alternating-projection–based solvers often suffer from instability under experimental noise, motivating smoother loss functions and momentum-based optimization schemes[42]. By comparison, the SAF + NAG solver maintains consistent intensity and sharpness throughout the sequence, with minimal frame-to-frame variation despite object motion. This enhanced temporal stability can be attributed to the combination of a smoothed amplitude-based loss, which suppresses noise-induced fluctuations, and Nesterov momentum, which dampens oscillatory behavior during iterative updates.

Together, these results demonstrate that SAF + NAG provides a temporally consistent reconstruction framework under motion, satisfying a key requirement for continuous holographic telepresence that extends beyond single-frame reconstruction accuracy.

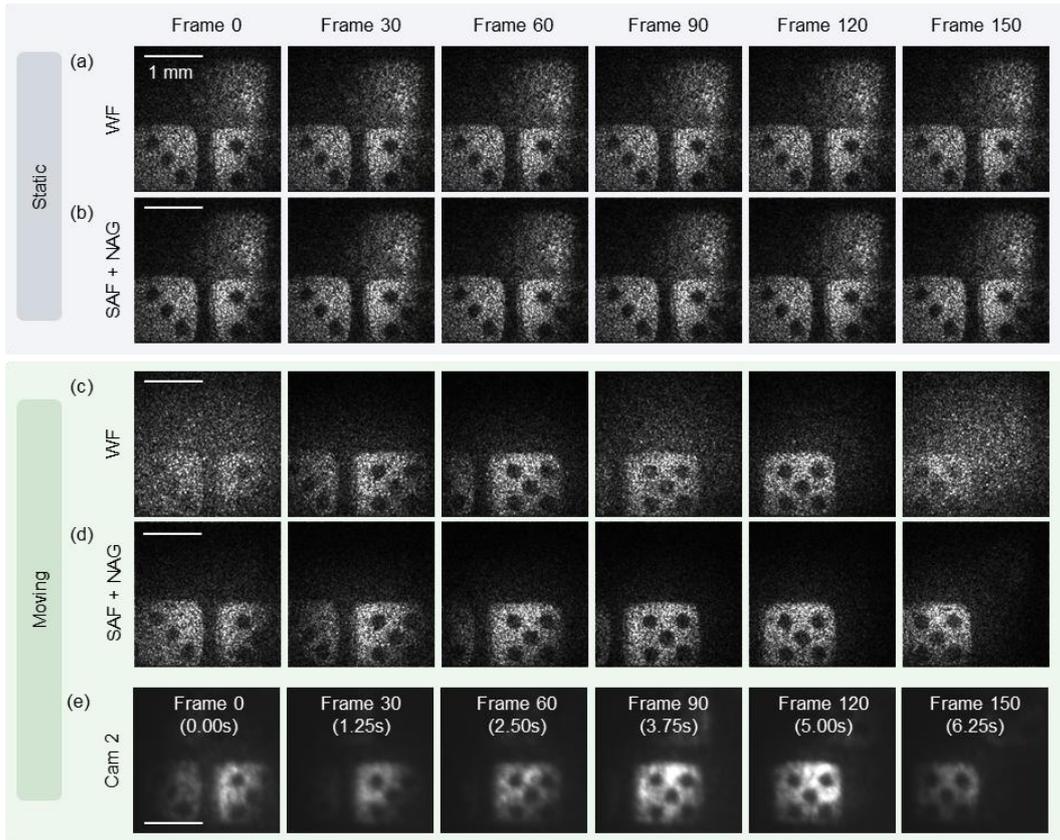

**Fig. 4 | Temporal robustness of wavefront reconstruction under static and dynamic scenes.** Representative frames extracted every 30 frames from a 150-frame video sequence reconstructed using either Wirtinger Flow (WF) or the proposed SAF + NAG solver. (a, b) Static scene consisting of three stationary dice. Both WF and SAF + NAG yield comparable reconstruction quality with minimal temporal fluctuation across frames. (c, d) Dynamic scene in which the dice undergo lateral translation. WF reconstructions exhibit pronounced frame-to-frame flicker and non-uniform brightness, particularly at the beginning and end of the sequence, indicating temporal instability under motion. In contrast, SAF + NAG maintains consistent intensity and sharpness throughout the sequence, demonstrating enhanced robustness for dynamic wavefront recovery. (e) Time-lapse recordings of a translating dice hologram optically replayed by the SLM and captured by the second camera, confirming temporally stable holographic projection. The camera frame rate was set to 24 frames per second. All reconstructed images in (a–d) are numerically generated from the recovered complex optical fields.

### 3.3 Video-rate telepresence

We next demonstrate video-rate holographic telepresence by directly projecting the reconstructed wavefronts onto the spatial light modulator (SLM) and recording the optically replayed fields with a second camera (Cam2) [Fig. 4(e)]. To streamline the system architecture, a single PC (PC1) was used for both acquiring raw speckle images from the camera and transmitting the computed phase patterns to the SLM. Wavefront reconstruction was performed on a dedicated remote server equipped with four RTX A6000 GPUs, with data exchanged over a 1-Gbps local network [Fig. 5(a)].

Under this configuration, the system achieved a sustained throughput of 28 frames per second (fps) with an end-to-end latency of 1.24 s, measured from speckle acquisition to SLM projection. Speckle images captured at the imaging stage were streamed to the server for multi-GPU wavefront reconstruction, and the resulting phase patterns were transmitted back to the display stage for holographic replay.

To quantitatively evaluate system performance and stability, we conducted a series of throughput and latency measurements. Figure 5(a) summarizes the end-to-end data flow, while Figs. 5(b) and 5(c) report the measured latency and output-rate stability, respectively. As shown in Fig. 5(b), the end-to-end latency remains nearly constant within the operational regime and increases sharply once the input frame rate exceeds the system throughput ($\approx$28 fps), indicating saturation of the reconstruction pipeline. At 28 fps, the measured latency was consistently 1.24 s. The total data transfer required for video-rate operation was approximately 896 Mbps (28 fps × 4 MB per frame), remaining below the available network bandwidth. This observation indicates that the dominant performance bottleneck arises from computational wavefront reconstruction, rather than network transmission.

Temporal stability was further assessed by monitoring the output frame rate under a constant input rate of 28 fps [Fig. 5(c)]. The output remained centered around 28 fps with a standard deviation of 2.64 fps, demonstrating stable streaming and holographic replay. Together, these measurements confirm sustained video-rate telepresence with predictable latency, validating the practical feasibility of the proposed wavefront-based telepresence pipeline.

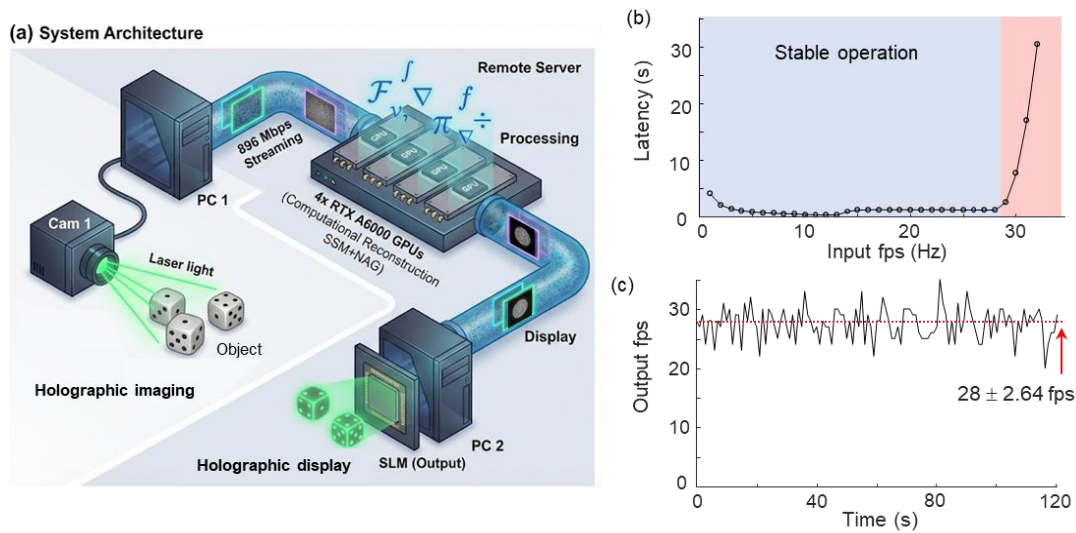

**Fig. 5 | System architecture and performance evaluation of video-rate holographic telepresence.** (a) End-to-end system architecture illustrating the data flow from single-shot speckle acquisition to holographic wavefront replay.

(b) End-to-end latency from speckle capture to SLM projection as a function of input frame rate. Each data point represents the median latency over repeated measurements. (c) Temporal stability of the output frame rate when the input frame rate is fixed at 28 frames per second.

## 4. Discussion

The results presented in this work demonstrate that physically faithful holographic telepresence can be achieved through direct wavefront measurement and replay, without relying on interferometric reference beams or geometry-based scene representations. By combining reference-free speckle-correlation imaging with single-shot SSM initialization and SAF + NAG refinement, the proposed framework enables stable recovery of complex optical wavefronts and their direct projection onto a phase-only spatial light modulator.

A key observation from the experimental results is that temporal consistency, rather than per-frame reconstruction accuracy alone, constitutes the critical requirement for holographic telepresence. While both WF and SAF + NAG provide comparable reconstructions for static scenes, only the SAF + NAG solver maintains stability under object motion, suppressing frame-to-frame flicker and brightness fluctuations.

System-level evaluation further indicates that the proposed architecture can sustain video-rate operation at 28 fps with predictable latency. Importantly, the measured end-to-end delay and throughput analysis reveal that the dominant performance bottleneck arises from computational wavefront reconstruction, rather than optical acquisition or network transmission. This separation clarifies that the current limitations are not fundamental to the optical architecture, but instead reflect the present computational implementation.

Several limitations remain in the current system. The reconstruction pipeline is computationally intensive, resulting in an end-to-end latency of approximately 1.24 s. In addition, the present implementation operates under monochromatic illumination and exhibits residual speckle artifacts associated with the use of a temporally coherent laser source. The numerical aperture of the display optics also constrains achievable resolution and viewing angles. Addressing these limitations will require advances in both computation and optics, including faster optimization strategies, color multiplexing schemes, higher-throughput optical designs, and improved speckle suppression techniques.

Beyond the specific system demonstrated here, these findings place the proposed approach within a broader context of emerging holographic communication technologies. Recent progress in computational holography and neural scene rendering has enabled impressive visual synthesis[19], yet such methods fundamentally rely on rendered approximations of optical propagation. In contrast, the present work emphasizes a measurement-driven paradigm, in which the optical wavefront itself is treated as the primary data object for capture, transmission, and display.

## 5. Conclusion

In conclusion, we have presented a reference-free holographic telepresence system that captures and replays complex optical wavefronts from a single intensity-only speckle measurement. The proposed framework achieves video-rate operation (≈28 fps) with modest communication bandwidth (<1 Gbps), while preserving physically consistent volumetric cues through direct wavefront propagation rather than geometric or view-synthesized approximations.

By integrating speckle-correlation scattering matrix reconstruction with SAF + NAG–based iterative refinement, the system enables stable holographic reconstruction and holographic replay under both static and dynamic conditions. Experimental validation confirms that single-shot wavefront acquisition, combined with physically grounded reconstruction and optical replay, provides a viable route toward practical holographic telepresence.

More broadly, this work suggests that measured holography can serve as a powerful complement to rendered holography, particularly for applications that demand physical fidelity, depth consistency, and propagation-accurate three-dimensional visualization. Looking forward, future holographic communication systems are likely to integrate data-driven inference with direct wavefront measurement, providing both computational efficiency and optical fidelity. The framework demonstrated here represents an important step toward such physically grounded holographic telepresence technologies.

## Acknowledgements

We are grateful for financial supports from National Research Foundation of Korea (RS-2024-00442348, 2022M3H4A1A02074314), Korea Institute for advancement of Technology (KIAT) (P0028463), the Korean Fund for Regenerative Medicine (KFRM) (21A0101L1-12), Samsung Research Funding Center of Samsung Electronics (SRFC-IT1401-08).

## Author contributions

All authors reviewed and commented on the manuscript. Chansuk Park, KyeoReh Lee, and YongKeun Park conceived the original idea, and YongKeun Park supervised the project. Chansuk Park and Chulmin Oh performed the measurements. Minwook Kim analyzed the measurement data and wrote the manuscript. Minwook Kim, Chansuk Park, and Chulmin Oh contributed equally to this work.

## Competing interests

The authors declare no conflicts of interest.